\documentclass[reprint,amsmath,amssymb,superscriptaddress,prl]{revtex4-2}
\usepackage{graphicx}
\usepackage{epstopdf}
\usepackage{dcolumn}
\usepackage{bm}
\usepackage{lineno}
\begin{document}

\title{Enhanced Nonlinear Frequency Conversion Bandwidth through Birefringence induced Mode Hybridization}

\author{Tingge Yuan}
\affiliation{State Key Laboratory of Advanced Optical Communication Systems and Networks, School of Physics and Astronomy, Shanghai Jiao Tong
University, 800 Dongchuan Road, Shanghai 200240, China
}

\author{Jiangwei Wu}
\affiliation{State Key Laboratory of Advanced Optical Communication Systems and Networks, School of Physics and Astronomy, Shanghai Jiao Tong
University, 800 Dongchuan Road, Shanghai 200240, China
}

\author{Xueyi Wang}
\affiliation{State Key Laboratory of Advanced Optical Communication Systems and Networks, School of Physics and Astronomy, Shanghai Jiao Tong
University, 800 Dongchuan Road, Shanghai 200240, China
}

\author{Hao Li}
\affiliation{State Key Laboratory of Advanced Optical Communication Systems and Networks, School of Physics and Astronomy, Shanghai Jiao Tong
University, 800 Dongchuan Road, Shanghai 200240, China
}

\author{Yuping Chen}
\email{ypchen@sjtu.edu.cn}
\affiliation{State Key Laboratory of Advanced Optical Communication Systems and Networks, School of Physics and Astronomy, Shanghai Jiao Tong
University, 800 Dongchuan Road, Shanghai 200240, China
}

\author{Xianfeng Chen}
\affiliation{State Key Laboratory of Advanced Optical Communication Systems and Networks, School of Physics and Astronomy, Shanghai Jiao Tong
University, 800 Dongchuan Road, Shanghai 200240, China
}
\affiliation{Collaborative Innovation Center of Light Manipulations and Applications, Shandong Normal University, Jinan 250358, China}
\date{\today}

\begin{abstract}
\noindent
On-chip quantum information network requires qubit transfer between different wavelengths while preserving quantum coherence and entanglement, which needs broadband up-conversion available. Herein, we demonstrate a mode-hybridization based broadband nonlinear frequency conversion on X-cut thin film lithium niobate. With the spontaneous quasi-phase matching and quasi group-velocity matching being simultaneously satisfied, broadband second harmonic generation with a 3-dB bandwidth up to 13 nm has been achieved in a micro-racetrack resonator. The same mechanism can work on the frequency conversion of the ultra-short pulse in the bent waveguide structure. This work will be beneficial to on-chip tunable frequency conversion and quantum light source generation on integrated photonic platforms, and further enable on-chip large-capacity multiplexing, multichannel optical information processing, and large quantum information networks. 
\end{abstract}

\maketitle
\begin{figure}[t]
\centering
 \includegraphics[width=\linewidth]{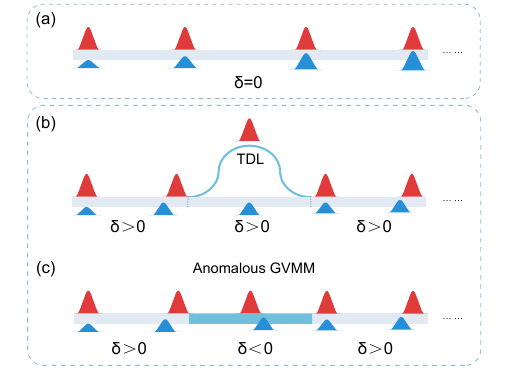}
	\caption{Schematics of the (a) direct GVM\cite{16Jankowski2019UltrabroadbandNO}, (b) QGVM realized by TDL\cite{39Xie2006NarrowlinewidthNO} and (c) anomalous GVMM in the phase-matched second harmonic generation process, where $\delta={V_{\rm{gSH}}}^{-1}-{V_{\rm{gFW}}}^{-1}$.}
	\label{fig1}
\end{figure}

\noindent Efficient second-order nonlinear process with widely-tunable pump bandwidth has always been the pursued goal owing to the extensive applications in wavelength division multiplexing network\cite{1Yoo1996WavelengthCT}, ultra-short pulse nonlinearity\cite{2Guo2022FemtojouleFA}, quantum key distribution\cite{3Adachi2006SimpleAE,4Leach2010QuantumCI,5Lo2014SecureQK}, and broadband single-photon source generation\cite{6Nasr2008UltrabroadbandBG,7Javid2021UltrabroadbandEP}. Generally, broad nonlinear bandwidth requires the phase-matching condition to be satisfied over a wide spectral range, which is equivalent to the simultaneous matching of both the group-velocity and phase-velocity of the interacting waves in the time domain\cite{7.1Yu2002BroadbandQS,7.2Zhang2008FlexibleWC,7.3Gong2010AllOW}. \\
\indent In the last decade, thin-film lithium niobate (TFLN) has emerged as an ideal platform to manipulate and investigate the nonlinear interaction in the wavelength-scale\cite{8Qi2020IntegratedLN,9Lin2020AdvancesIO,10Boes2023LithiumNP,ma2020ultrabright, chen2021photon}, where the compact optical structures could not only enable the huge enhancement of the conversion efficiency by tightly confining the optical filed\cite{11Wang2018UltrahighefficiencyWC,12Chen2019EfficientPF,13Zhao2020ShallowetchedTL}, but also provide degrees of freedom to tailor the group-velocity, as well as the group-velocity dispersion through the structural geometry\cite{14Wang2018MonolithicLN}. As a result, broadband second harmonic generation (SHG) under the waveguide configuration on TFLN has been extensively studied in recent years\cite{15Ge2018BroadbandQM,16Jankowski2019UltrabroadbandNO,17Mishra:22,18Javid2021UltrabroadbandEP}. Using quasi-phase matching (QPM) technique\cite{19Zhu1997QuasiphasematchedTG,20Paul2003QuasiphasematchedGO}, broadband SHG bandwidth over hundreds of manometers has been demonstrated in the dispersion-engineered periodically poled ridged waveguide on TFLN\cite{16Jankowski2019UltrabroadbandNO}, where the fundamental wave (FW) and second harmonic wave (SH) have the same group velocity and stable overlap over the whole propagation distance, as the diagram in Fig.\ref{fig1} (a) has illustrated. Other methods that introduce the non-uniformity to broaden the bandwidth have also been reported\cite{21Chen2015HighEfficiencyBH,22Huang2022WidefieldMS,23Wang2017MetasurfaceassistedPS}, while these conversion efficiencies are relatively low due to the limited interaction length.\\ 
\indent In addition to direct GVM by dispersion engineering or material nature, another alternative method of quasi-group velocity matching (QGVM) has been used to achieve the broadband nonlinear frequency conversion. This idea was first proposed by M. M. Fejer in 2004\cite{38Huang2004QuasigroupvelocityMU} (see Fig.\ref{fig1}(b)), which introduced a wavelength-selected time-delay line (TDL) in periodically poled bulk LN\cite{39Xie2006NarrowlinewidthNO}, and latter applied to fiber optics\cite{40Mao2021SynchronizedMS,41Lourdesamy2021SpectrallyPP,42Cui12022DichromaticM}. In this scheme, group velocities of interacting waves are no longer required to be always matched along the propagation direction, but rather the structure geometry is required to satisfy a certain ratio depending on the accumulated group-velocity mismatch. However, TDL-based devices typically suffer from a relatively large footprint, which puts forward an urgent need to realize the chip-scale integration for QGVM-based broadband nonlinearity.\\ 
\indent Recently, anisotropic properties of the X-cut TFLN have been widely studied and applied to the electro-optic modulation\cite{wang2018integrated,zhang2019broadband}, polarization manipulation\cite{han2022mode,9095209}, sensing resolution enhancement\cite{wang2024enhanced}, and poling-free QPM with the spontaneously inverted nonlinear coefficient\cite{lin2019broadband,352023ChipscaleSQ}. To tailor the dispersion as well as the group velocity, such anisotropy also provides a potential path through the birefringence-induced mode-hybridization\cite{36Pan2019FundamentalMH,37Liu2022ModalAO}, which can introduce the anomalous GVMM and compensate the temporal walk-off between FW and SH waves in a single waveguide, as Fig.\ref{fig1}(c) has illustrated. Based on the two routes, this paper demonstrates the greatly improved nonlinear bandwidth in the micro-nanophotonic structures on X-cut TFLN. Our experimental results show that broadband SHG can be found in a racetrack-resonator or bent waveguide with a specific geometrical structure, in which the spontaneous quasi phase-matching (SQPM) and QGVM are proven to be simultaneously satisfied in our simulation. Our study will open a new avenue towards the broadband nonlinear functional devices on the anisotropic materials based on-chip photonic platform.\\
\begin{figure*}[t]
	\centering
 \includegraphics[width=\linewidth]{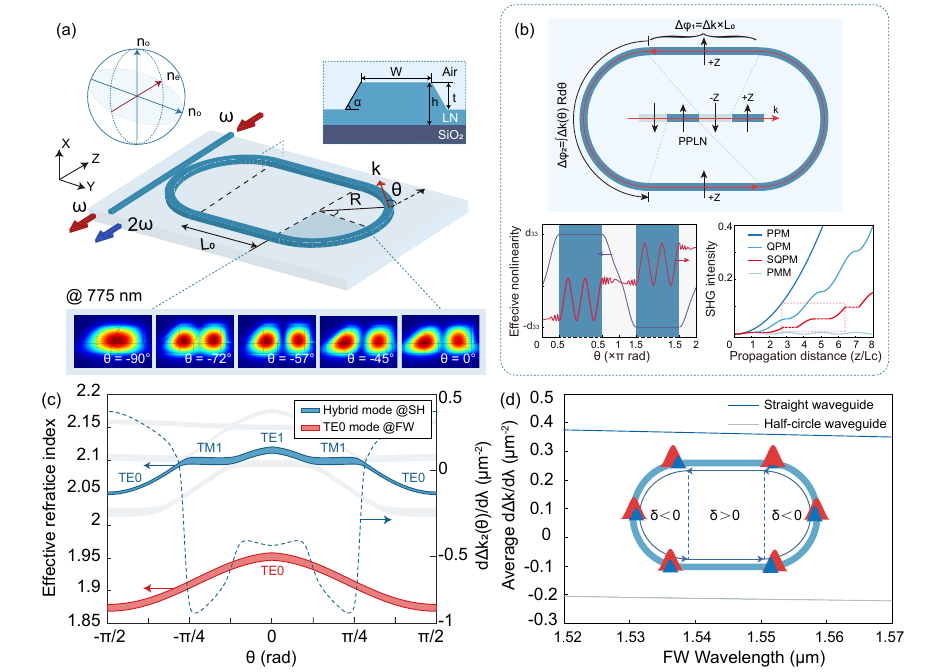}
	\caption{(a) Schematic of the birefringent racetrack resonator on X-cut TFLN, where SH-band light experiences a mode-hybridization in the half-circle waveguide. (b) Principle of SQPM. Inset: varying SQPM SHG intensity with the periodically inverted efficient nonlinear coefficient ($m=5$), and a comparison among the SHG processes under the perfect phase-matching (PPM), QPM, SQPM, and phase mismatching (PMM). (c) Effective refractive indices of the hybrid mode in SH-band and TE0 mode in FW-band in the half-circle waveguide, and the vector mismatch dispersion between them. (d) Average vector mismatch dispersion versus different FW wavelength, which is positive in the straight waveguide and negative in the half-circle waveguide.}
	\label{fig2}
\end{figure*}
\indent A schematic of the typical structure proposed in this work is shown in Fig.\ref{fig2}(a). Here a general SHG process is considered in a micro-racetrack resonator on X-cut TFLN, which has a straight section length of $L_{0}$ and the half-circle radius of $R$. The straight waveguide is aligned with the Y axis. As the TE-polarized fundamental wavelength (FW) and second harmonic (SH) light circulate in the racetrack resonator, corresponding effective nonlinear coefficient $d_{\rm{eff}}$ varies with the direction of the wave vector $k$ in the relationship of:
\begin{equation}
d_{\rm{eff}}(\theta)=-d_{\rm{22}}{{\cos }^3}\theta+d_{\rm{31}}{{\cos }^2}\theta{\sin}\theta+d_{\rm{33}}{{\sin }^3}\theta.
\label{eq:SQPM deff}   
\end{equation}
where $\theta$ is the propagation azimuthal angle. From the above equation we can see that $d_{\rm{eff}}$ oscillates between the $d_{33}$ and $-d_{33}$ in a $2\pi$-period of $\theta$, and it reaches the maximum and minimum values in two straight waveguides respectively. The SHG process also depends on the phase mismatch between the FW and SH light. In the different sections of the racetrack resonator, the accumulated phase-mismatch can be calculated as follows:
\begin{equation}
\begin{aligned}
&\Delta \phi_{1}=\Delta k_{1} L_{0},\\
&\Delta \phi_{2}=\int_{0}^{\pi} \Delta k_{2}(\theta) R d\theta=\Delta k_{2}^{'}\pi R.
\end{aligned}
\label{dphi define}
\end{equation}
where $\Delta k_{1}$ and $\Delta k_{2}(\theta)$ are the vector-mismatch in the straight waveguide and the birefringent half-circle waveguide, respectively, which have the expression of the $\Delta k_{1(2)}=4\pi \Delta n_{1(2)}/\lambda$ with the FW wavelength $\lambda$ and the effective refractive index difference $\Delta n_{1(2)}$ between the FW and SH. For simplicity, we use $\Delta k_{2}^{'}$ as a reduced form of $\Delta k_{2}(\theta)$ in the following analysis, which is defined by $\Delta k_{2}^{'}=\frac{1}{\pi}\int_{0}^{\pi} \Delta k_{2}(\theta) d\theta$. Analogous to the common QPM process in the periodically poled lithium niobate, the realization of the perfect SQPM requires $\Delta \phi_{1} = m\pi$ and $\Delta \phi_{2} = 2N\pi$, where $m$ is an odd number and $N$ is an integer, corresponding to the QPM condition and  continuation condition, respectively. In this case, the SH intensity will grow continuously in the poling-free racetrack resonator, as shown by the inset in Fig.\ref{fig2}(b). Furthermore, at the cost of some conversion efficiency, the above requirements can be further relaxed to the condition of
\begin{equation}
 \Delta \phi_{1}+\Delta \phi_{2}=(2N+m)\pi, \quad(\Delta\phi_{1}\bmod{2\pi}\neq 0)
 \label{sqpm}
\end{equation}
which means that the slight shift from the $m\pi$ in the $\Delta \phi_{1}$ can be compensated by that in the $\Delta \phi_{2}$, so that an overall phase relationship between the FW and SH in a period remains unchanged. Compared to the other phase-matching methods, the conversion efficiency of the SQPM depends mainly on the structure size, and is therefore usually lower than that of the common PPLN-based QPM.\\
\indent Based on Eq.\ref{sqpm}, we have the broadband SQPM condition which requires at least the first-order derivative of $\Delta \phi_{1}+\Delta \phi_{2}$ to be equal to 0 at the central FW wavelength $\lambda_{0}$, which satisfies SQPM. Combining Eq.\ref{dphi define} and \ref{sqpm}, it can be derived to an equation related to the racetrack structure and dispersion, in the form of:
\begin{equation}
\centering
    L_{0}(\frac{d \Delta k_{1}}{d \lambda})_{\lambda_{0}}=-\pi R(\frac{d {\Delta k_{2}}^{'}}{d \lambda})_{\lambda_{0}}.
        \label{eq2}
\end{equation}
\begin{figure*}[t]
	\centering
 \includegraphics[width=\linewidth]{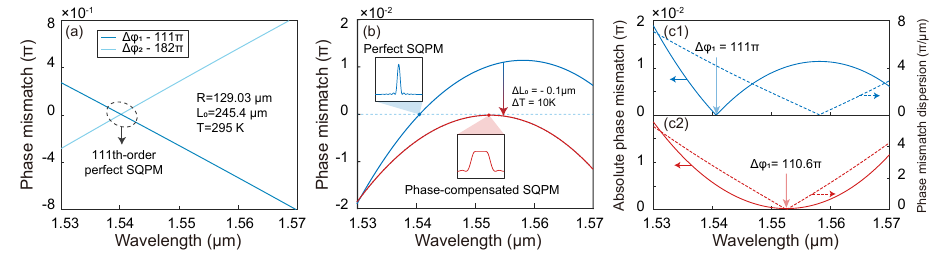}
	\caption{(a) Calculated phase mismatch $\Delta \phi_{1} $ in the straight waveguide and phase mismatch $\Delta \phi_{2}$ in the half-circle waveguide of a perfect 111-order SQPM racetrack resonator, and (b) their summation, presented by $\Delta \phi_{1}+\Delta \phi_{2}-293 \pi$. A detailed comparison between (c1) the perfect narrowband SQPM and (c2) a phase-compensated broadband SQPM. The solid and dashed lines denote the absolute phase-mismatch summation $|\Delta \phi_{1}+\Delta \phi_{2}-293 \pi|$ and its dispersion $|d(\Delta \phi_{1}+\Delta \phi_{2})/d\lambda|$, respectively.}
	\label{fig3}
\end{figure*}
\indent Generally, it's difficult to fulfil Eq.\ref{eq2} in an isotropic platform because the sign of $d \Delta k /d \lambda$ is hard to change due to the relatively consistent mode transition between the straight and half-circle waveguides. However, for a birefringent racetrack resonator, the potential mode-hybridization in the half-circle waveguide could provide another degree of freedom to manipulate the modal dispersion. For example, in a ridge waveguide with a top width of 1 $\mu$m and a height of 0.38 $\mu$m, the FW light with a wavelength of 1550 nm will maintain the single mode condition during the intracavity propagation, while the SH light with the wavelength of 775 nm will experience a strong mode-hybridization in the half-circle waveguide, as shown in the Fig.\ref{fig2}(a). Therefore, its modal dispersion, as well as $\rm{d} {\Delta k_{2}}^{'}/\rm{d} \lambda$ can be significantly changed compared to the $\rm{d} {\Delta k_{1}}/\rm{d} \lambda$ in the straight waveguide. \\
\indent Specifically, we have simulated the effective refractive indices of all possible modes in the SH and FW bands in the half-circle waveguide on the X-cut TFLN. Conformal transformation was used to deal with the effect of the waveguide bending.  In the SH band, the imported TE0 mode in the straight waveguide gradually transforms to the TM1 mode and then to the TE1 mode, as the propagation direction has been rotated by 90 degrees in the quarter-circle waveguide. In Fig.\ref{fig2}(c), the width of the blue and red lines represent the change of indices in the FW wavelength range from 1530 nm to 1570 nm, from which it is obvious to see the different dispersion property at three stages of SH. The dashed line shows the $\theta$ dependent d$\Delta k_{2}$/d$\lambda$ at the FW wavelength of 1550 nm in the half-circle waveguide. When both the FW and SH are in the TE0 modes, d$\Delta k_{2}$/d$\lambda$ has a positive values as in the straight waveguide; while when the SH transfers to the higher-order modes, the sign of d$\Delta k_{2}$/d$\lambda$ rapidly changes to the negative. The average vector-mismatch dispersion in the straight and half-circle waveguides has been calculated versus the FW wavelength in Fig.\ref{fig2}(d), which shows an opposite sign over a broadband from 1520 nm to 1570 nm. From a time domain point of view, the temporal walk-off accumulated in the straight waveguide between the interacting pulses in the FW and SH bands can be reduced or even compensated if the racetrack geometry satisfies Eq.\ref{eq2}, which means that these two pulses will overlap again when they enter the next straight waveguide.\\
\begin{figure*}[t]
 \includegraphics[width=\linewidth]{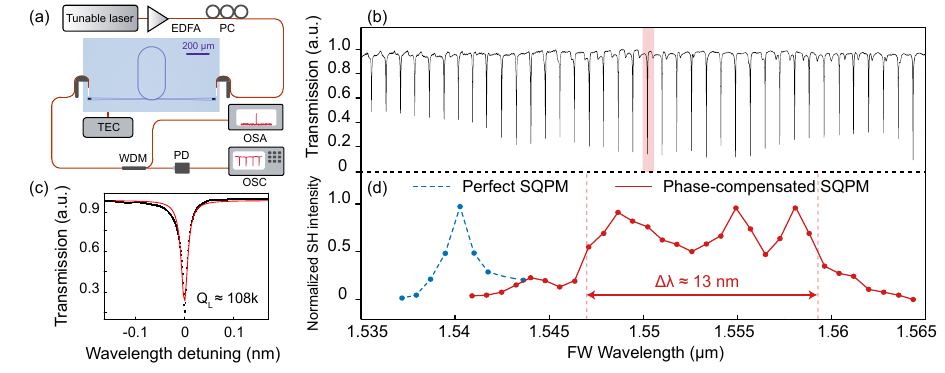}
	\caption{(a) Experimental setup.  EDFA: erbium-doped optical fibre amplifier. PC: polarization controller. TEC: thermal electronic cooler. WDM: wavelength division multiplexer. OSA:optical spectrum analyser. PD: photodetector. OSC: oscilloscope.(b) Transmission spectrum of the SQPM racetrack resonator in C-band,  and (c) Lorentzian fitting of the marked resonance dip. (d) SHG intensity obtained at each FW resonance mode.}
	\label{fig4}
\end{figure*}
\indent To experimentally demonstrate the mode hybridization induced broadband SQPM SHG, we have first designed a dual-resonant 111th-order perfect SQPM racetrack resonator with a straight waveguide length $L_{0}$ of 245.4 $\mu m$ and a half-circle radius $R$ of 129.03 $\mu m$, corresponding to $\Delta \phi_{1}=111\pi$ and $\Delta \phi_{2}=2\times 91\pi$ at the FW wavelength of 1540 nm. Details of the design process can be found in Ref\cite{352023ChipscaleSQ}. It is worth noting that the choice of geometric parameters is mainly determined by the fabrication process. By improving the machining process or using the deep-etching to suppress bending loss, the size of the micro-cavity can be further reduced if the above conditions are met. In this work we focus on the dispersion property of this structure as shown in Fig.\ref{fig3}(a). For the designed $R$ and $L_{0}$, $\Delta \phi_{1}$ decreases for the longer FW wavelength and $\Delta \phi_{2}$ vice versa, both of which have a close absolute values of the slope. Furthermore, we calculate the phase-mismatch summation in Fig.\ref{fig3}(b), and find it takes an extremum ($\rm{d}$$(\Delta \phi_{1}+\Delta \phi_{2})/$$\rm{d}$$\lambda=0$) around the FW wavelength of 1560 nm, while the QPM condition ($\Delta \phi_{1}+\Delta \phi_{2}=293\pi$) is fulfilled at that of 1540 nm. By slightly tuning the temperature or the geometric parameter, the SQPM and QGVM conditions can be realized simultaneously. For example, as the temperature has risen by 10 K, and the $L_{0}$ has been reduced by 100 nm, the zero points of $\Delta \phi_{1}+\Delta \phi_{2}-293\pi$ and its first-order dispersion may shift and eventually meet at the FW wavelength of about 1552 nm. A detailed comparison is shown in the Fig.\ref{fig3}(c). At the central FW wavelength of the perfect SQPM condition, $\Delta \phi_{1}$ is equal to $111\pi$, indicating a theoretically highest SH intensity; while it is shifted to $110.6\pi$ under the broadband SQPM condition, the overall SH intensity will be slightly lower. However, since the absolute change of $\Delta \phi_{1}$ is always within $\pi$, and in most cases less than $\pi/2$ as the FW wavelength is scanned from 1530 nm to 1570 nm, the conversion efficiency is not affected too much.\\
\begin{figure*}[t]
 \includegraphics[width=\linewidth]{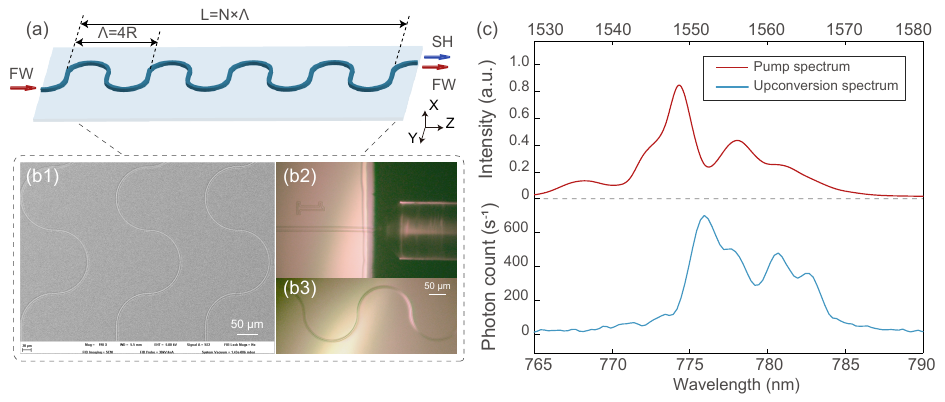}
	\caption{(a) Schematics of the SQPM bent-waveguide on the X-cut TFLN. (b) Scanning electron microscopic image of the first-order SQPM waveguide.(c) Pump light spectrum and measured nonlinear upconversion spectrum, corresponding to a 3-dB bandwidth of about 16 nm.}
	\label{fig5}
\end{figure*}
\indent In our experiment, 111th-order perfect and phase-compensated SQPM racetrack resonators have been fabricated and characterized, and the fabrication process can be found in Ref\cite{352023ChipscaleSQ}. A pulley-type bus waveguide is used for on-chip coupling of the resonator, which has a centre angle of 30 degrees, top width of 0.8 $\mu$m, and gap of 0.6 $\mu$m, respectively. Fig.\ref{fig4}(a) shows the experimental setup. A pair of grating couplers is used to couple the light in and out of the chip. The transmission spectrum of the FW light is recorded by an OSC under the small signal condition, and that of the shorter racetrack resonator is shown in Fig.\ref{fig4} (b). A typical TE0 mode around the wavelength of 1550 nm is zoomed in and fitted by the Lorentzian function, which shows an intrinsic Q factor of 155k. Measured SH intensities in two resonators at different resonant modes of FW are shown in Fig.\ref{fig4}(d), from which a relatively large SHG spectrum from 1540 nm to 1565 nm can be observed with a 3-dB bandwidth of about 13 nm under phase-compensated SQPM. Compared with the SH bandwidth we obtained in the perfect SQPM resonator, the bandwidth is significantly enhanced over ten-fold. For the broadband SHG, the highest on-chip SH power of about 100 nW is measured at the FW wavelength of 1555 nm under an on-chip pump power of 20 mW. The relatively low conversion efficiency can be attributed to the large SQPM order-number used here. By exploring the lower-order SQPM (see the supplemental materials) or periodically poling the straight waveguide, broad bandwidth and an improved conversion efficiency can be achieved simultaneously. Besides, we find the SH intensity has a relatively large fluctuation with the FW wavelength from 1548 nm to 1558 nm, which can be attributed to the different coupling condition and resonant mode mismatch.\\
\indent Furthermore, we show that the mode hybridisation induced QGVM can be combined with the SQPM in the bent waveguide configuration, as the schematic shown in Fig.\ref{fig5}(a). Compared with the SQPM-based racetrack resonator, the SQPM bent waveguide breaks the restriction of resonance, thus can be applied to generate the SH light in a continuous broadband spectrum instead of a series of discrete modes. By selecting different hybrid mode in the SH band, we have designed the first-order SQPM bent waveguide with the QGVM satisfied at the same time, and a detailed theoretical analysis can be found in the Supplemental Materials. As shown in the Fig.\ref{fig5}(b), here the half-circle radius $R$ is 75 $\mu$m, the straight section length $L_{0}$ is about 2.2 $\mu$m, corresponding to the coherent length of the SHG process involved in TE0 modes at the FW wavelength of 1550 nm. The total length $L$ of the bent waveguide in the Z-direction is about 15 mm, including 50 SQPM cycles. A fibre lens is used for coupling at the edge of the waveguide. The pump light for the FW is provided by a C-band femtosecond laser with a repetition frequency of 50 MHz and a pulse duration of about 500 fs. The spectrum of the pump light is shown in Fig\ref{fig5}(c) with the red curve. In our experiment, with an on-chip average pump power of about 5.7 mW, a broadband upconversion spectrum at the 775 nm band is measured by the OSA at the output port as the blue curve shown in Fig\ref{fig5}(c), which means that at least 16 nm 3dB bandwidth has been achieved in the SQPM bent waveguide structure. According to the theoretical prediction, this bandwidth could be up to about 50 nm. It should be noted that the conversion efficiency of the bent waveguide in our experiment is mainly limited by the high insertion loss ($>$50 dB). By optimising the structure parameters or improving the fabrication process, the conversion efficiency could be further increased.\\
\indent  In conclusion, we have demonstrated a new approach to achieve the QGVM in the racetrack resonator or bent waveguide on X-cut TFLN. Based on the birefringence-induced mode transition of the SH light, the average group velocity mismatch in the SHG process can be flexibly tuned with the proper design of the racetrack resonator geometry. A tenfold increase in the intracavity SHG bandwidth and frequency conversion of the femtosecond laser pulse was experimentally observed. With further dispersion engineering and optimization of the structure, on-chip nonlinear frequency conversion between the ultrashort optical pulses and even the quantum states can be expected.\\
\begin{acknowledgments}
This work was supported by the National Natural Science Foundation of China (Grant Nos. 12134009), the National Key R\&D Program of China (Grant Nos. 2019YFB2203501), Shanghai Municipal Science and Technology Major Project (2019SHZDZX01-ZX06), and SJTU No. 21X010200828.\\
The authors would like to thank Prof. Wenjie Wan for his helpful discussion of our work. \\
*The authors declare no conflicts of interest.
\nocite{*}
\end{acknowledgments}


\begin{thebibliography}{47}%
\makeatletter
\providecommand \@ifxundefined [1]{%
 \@ifx{#1\undefined}
}%
\providecommand \@ifnum [1]{%
 \ifnum #1\expandafter \@firstoftwo
 \else \expandafter \@secondoftwo
 \fi
}%
\providecommand \@ifx [1]{%
 \ifx #1\expandafter \@firstoftwo
 \else \ex9andafter \@secondoftwo
 \fi
}%
\providecommand \natexlab [1]{#1}%
\providecommand \enquote  [1]{``#1''}%
\providecommand \bibnamefont  [1]{#1}%
\providecommand \bibfnamefont [1]{#1}%
\providecommand \citenamefont [1]{#1}%
\providecommand \href@noop [0]{\@secondoftwo}%
\providecommand \href [0]{\begingroup \@sanitize@url \@href}%
\providecommand \@href[1]{\@@startlink{#1}\@@href}%
\providecommand \@@href[1]{\endgroup#1\@@endlink}%
\providecommand \@sanitize@url [0]{\catcode `\\12\catcode `\$12\catcode `\&12\catcode `\#12\catcode `\^12\catcode `\_12\catcode `\%12\relax}%
\providecommand \@@startlink[1]{}%
\providecommand \@@endlink[0]{}%
\providecommand \url  [0]{\begingroup\@sanitize@url \@url }%
\providecommand \@url [1]{\endgroup\@href {#1}{\urlprefix }}%
\providecommand \urlprefix  [0]{URL }%
\providecommand \Eprint [0]{\href }%
\providecommand \doibase [0]{https://doi.org/}%
\providecommand \selectlanguage [0]{\@gobble}%
\providecommand \bibinfo  [0]{\@secondoftwo}%
\providecommand \bibfield  [0]{\@secondoftwo}%
\providecommand \translation [1]{[#1]}%
\providecommand \BibitemOpen [0]{}%
\providecommand \bibitemStop [0]{}%
\providecommand \bibitemNoStop [0]{.\EOS\space}%
\providecommand \EOS [0]{\spacefactor3000\relax}%
\providecommand \BibitemShut  [1]{\csname bibitem#1\endcsname}%
\let\auto@bib@innerbib\@empty
\bibitem [{\citenamefont {Yoo}(1996)}]{1Yoo1996WavelengthCT}%
  \BibitemOpen
  \bibfield  {author} {\bibinfo {author} {\bibfnamefont {S.~J.~B.}\ \bibnamefont {Yoo}},\ }\href {https://api.semanticscholar.org/CorpusID:110731121} {\bibfield  {journal} {\bibinfo  {journal} {Journal of Lightwave Technology}\ }\textbf {\bibinfo {volume} {14}},\ \bibinfo {pages} {955} (\bibinfo {year} {1996})}\BibitemShut {NoStop}%
\bibitem [{\citenamefont {Guo}\ \emph {et~al.}(2022)\citenamefont {Guo}, \citenamefont {Sekine}, \citenamefont {Ledezma}, \citenamefont {Nehra}, \citenamefont {Dean}, \citenamefont {Roy}, \citenamefont {Gray}, \citenamefont {Jahani},\ and\ \citenamefont {Marandi}}]{2Guo2022FemtojouleFA}%
  \BibitemOpen
  \bibfield  {author} {\bibinfo {author} {\bibfnamefont {Q.}~\bibnamefont {Guo}}, \bibinfo {author} {\bibfnamefont {R.}~\bibnamefont {Sekine}}, \bibinfo {author} {\bibfnamefont {L.}~\bibnamefont {Ledezma}}, \bibinfo {author} {\bibfnamefont {R.}~\bibnamefont {Nehra}}, \bibinfo {author} {\bibfnamefont {D.~J.}\ \bibnamefont {Dean}}, \bibinfo {author} {\bibfnamefont {A.}~\bibnamefont {Roy}}, \bibinfo {author} {\bibfnamefont {R.~M.}\ \bibnamefont {Gray}}, \bibinfo {author} {\bibfnamefont {S.}~\bibnamefont {Jahani}},\ and\ \bibinfo {author} {\bibfnamefont {A.}~\bibnamefont {Marandi}},\ }\href {https://api.semanticscholar.org/CorpusID:256703969} {\bibfield  {journal} {\bibinfo  {journal} {Nature Photonics}\ }\textbf {\bibinfo {volume} {16}},\ \bibinfo {pages} {625 } (\bibinfo {year} {2022})}\BibitemShut {NoStop}%
\bibitem [{\citenamefont {Adachi}\ \emph {et~al.}(2007)\citenamefont {Adachi}, \citenamefont {Yamamoto}, \citenamefont {Koashi},\ and\ \citenamefont {Imoto}}]{3Adachi2006SimpleAE}%
  \BibitemOpen
  \bibfield  {author} {\bibinfo {author} {\bibfnamefont {Y.}~\bibnamefont {Adachi}}, \bibinfo {author} {\bibfnamefont {T.}~\bibnamefont {Yamamoto}}, \bibinfo {author} {\bibfnamefont {M.}~\bibnamefont {Koashi}},\ and\ \bibinfo {author} {\bibfnamefont {N.}~\bibnamefont {Imoto}},\ }\href {https://api.semanticscholar.org/CorpusID:38698156} {\bibfield  {journal} {\bibinfo  {journal} {Physical review letters}\ }\textbf {\bibinfo {volume} {99 18}},\ \bibinfo {pages} {180503} (\bibinfo {year} {2007})}\BibitemShut {NoStop}%
\bibitem [{\citenamefont {Leach}\ \emph {et~al.}(2010)\citenamefont {Leach}, \citenamefont {Jack}, \citenamefont {Romero}, \citenamefont {Jha}, \citenamefont {Yao}, \citenamefont {Franke-Arnold}, \citenamefont {Ireland}, \citenamefont {Boyd}, \citenamefont {Barnett},\ and\ \citenamefont {Padgett}}]{4Leach2010QuantumCI}%
  \BibitemOpen
  \bibfield  {author} {\bibinfo {author} {\bibfnamefont {J.}~\bibnamefont {Leach}}, \bibinfo {author} {\bibfnamefont {B.}~\bibnamefont {Jack}}, \bibinfo {author} {\bibfnamefont {J.}~\bibnamefont {Romero}}, \bibinfo {author} {\bibfnamefont {A.~K.}\ \bibnamefont {Jha}}, \bibinfo {author} {\bibfnamefont {A.~M.}\ \bibnamefont {Yao}}, \bibinfo {author} {\bibfnamefont {S.}~\bibnamefont {Franke-Arnold}}, \bibinfo {author} {\bibfnamefont {D.}~\bibnamefont {Ireland}}, \bibinfo {author} {\bibfnamefont {R.~W.}\ \bibnamefont {Boyd}}, \bibinfo {author} {\bibfnamefont {S.~M.}\ \bibnamefont {Barnett}},\ and\ \bibinfo {author} {\bibfnamefont {M.~J.}\ \bibnamefont {Padgett}},\ }\href@noop {} {\bibfield  {journal} {\bibinfo  {journal} {Science}\ }\textbf {\bibinfo {volume} {329}},\ \bibinfo {pages} {662 } (\bibinfo {year} {2010})}\BibitemShut {NoStop}%
\bibitem [{\citenamefont {Lo}\ \emph {et~al.}(2014)\citenamefont {Lo}, \citenamefont {Curty},\ and\ \citenamefont {Tamaki}}]{5Lo2014SecureQK}%
  \BibitemOpen
  \bibfield  {author} {\bibinfo {author} {\bibfnamefont {H.-K.}\ \bibnamefont {Lo}}, \bibinfo {author} {\bibfnamefont {M.}~\bibnamefont {Curty}},\ and\ \bibinfo {author} {\bibfnamefont {K.}~\bibnamefont {Tamaki}},\ }\href {https://api.semanticscholar.org/CorpusID:119105614} {\bibfield  {journal} {\bibinfo  {journal} {Nature Photonics}\ }\textbf {\bibinfo {volume} {8}},\ \bibinfo {pages} {595 } (\bibinfo {year} {2014})}\BibitemShut {NoStop}%
\bibitem [{\citenamefont {Nasr}\ \emph {et~al.}(2008)\citenamefont {Nasr}, \citenamefont {Carrasco}, \citenamefont {Saleh}, \citenamefont {Sergienko}, \citenamefont {Teich}, \citenamefont {Torres}, \citenamefont {Torner}, \citenamefont {Hum},\ and\ \citenamefont {Fejer}}]{6Nasr2008UltrabroadbandBG}%
  \BibitemOpen
  \bibfield  {author} {\bibinfo {author} {\bibfnamefont {M.~B.}\ \bibnamefont {Nasr}}, \bibinfo {author} {\bibfnamefont {S.}~\bibnamefont {Carrasco}}, \bibinfo {author} {\bibfnamefont {B.~E.~A.}\ \bibnamefont {Saleh}}, \bibinfo {author} {\bibfnamefont {A.~V.}\ \bibnamefont {Sergienko}}, \bibinfo {author} {\bibfnamefont {M.~C.}\ \bibnamefont {Teich}}, \bibinfo {author} {\bibfnamefont {J.~P.}\ \bibnamefont {Torres}}, \bibinfo {author} {\bibfnamefont {L.}~\bibnamefont {Torner}}, \bibinfo {author} {\bibfnamefont {D.~S.}\ \bibnamefont {Hum}},\ and\ \bibinfo {author} {\bibfnamefont {M.~M.}\ \bibnamefont {Fejer}},\ }\href {https://api.semanticscholar.org/CorpusID:1461313} {\bibfield  {journal} {\bibinfo  {journal} {Physical review letters}\ }\textbf {\bibinfo {volume} {100 18}},\ \bibinfo {pages} {183601} (\bibinfo {year} {2008})}\BibitemShut {NoStop}%
\bibitem [{\citenamefont {Javid}\ \emph {et~al.}(2021{\natexlab{a}})\citenamefont {Javid}, \citenamefont {Ling}, \citenamefont {Staffa}, \citenamefont {Li}, \citenamefont {He},\ and\ \citenamefont {Lin}}]{7Javid2021UltrabroadbandEP}%
  \BibitemOpen
  \bibfield  {author} {\bibinfo {author} {\bibfnamefont {U.~A.}\ \bibnamefont {Javid}}, \bibinfo {author} {\bibfnamefont {J.}~\bibnamefont {Ling}}, \bibinfo {author} {\bibfnamefont {J.}~\bibnamefont {Staffa}}, \bibinfo {author} {\bibfnamefont {M.}~\bibnamefont {Li}}, \bibinfo {author} {\bibfnamefont {Y.}~\bibnamefont {He}},\ and\ \bibinfo {author} {\bibfnamefont {Q.}~\bibnamefont {Lin}},\ }\href {https://api.semanticscholar.org/CorpusID:231592825} {\bibfield  {journal} {\bibinfo  {journal} {Physical review letters}\ }\textbf {\bibinfo {volume} {127 18}},\ \bibinfo {pages} {183601} (\bibinfo {year} {2021}{\natexlab{a}})}\BibitemShut {NoStop}%
\bibitem [{\citenamefont {Yu}\ \emph {et~al.}(2002)\citenamefont {Yu}, \citenamefont {Ro}, \citenamefont {Cha}, \citenamefont {Kurimura},\ and\ \citenamefont {Taira}}]{7.1Yu2002BroadbandQS}%
  \BibitemOpen
  \bibfield  {author} {\bibinfo {author} {\bibfnamefont {N.~E.}\ \bibnamefont {Yu}}, \bibinfo {author} {\bibfnamefont {J.~H.}\ \bibnamefont {Ro}}, \bibinfo {author} {\bibfnamefont {M.}~\bibnamefont {Cha}}, \bibinfo {author} {\bibfnamefont {S.}~\bibnamefont {Kurimura}},\ and\ \bibinfo {author} {\bibfnamefont {T.}~\bibnamefont {Taira}},\ }\href {https://api.semanticscholar.org/CorpusID:27345139} {\bibfield  {journal} {\bibinfo  {journal} {Optics letters}\ }\textbf {\bibinfo {volume} {27 12}},\ \bibinfo {pages} {1046} (\bibinfo {year} {2002})}\BibitemShut {NoStop}%
\bibitem [{\citenamefont {Zhang}\ \emph {et~al.}(2008)\citenamefont {Zhang}, \citenamefont {Chen}, \citenamefont {Lu},\ and\ \citenamefont {Chen}}]{7.2Zhang2008FlexibleWC}%
  \BibitemOpen
  \bibfield  {author} {\bibinfo {author} {\bibfnamefont {J.}~\bibnamefont {Zhang}}, \bibinfo {author} {\bibfnamefont {Y.}~\bibnamefont {Chen}}, \bibinfo {author} {\bibfnamefont {F.}~\bibnamefont {Lu}},\ and\ \bibinfo {author} {\bibfnamefont {X.}~\bibnamefont {Chen}},\ }\href {https://api.semanticscholar.org/CorpusID:1562240} {\bibfield  {journal} {\bibinfo  {journal} {Optics express}\ }\textbf {\bibinfo {volume} {16 10}},\ \bibinfo {pages} {6957} (\bibinfo {year} {2008})}\BibitemShut {NoStop}%
\bibitem [{\citenamefont {Gong}\ \emph {et~al.}(2010)\citenamefont {Gong}, \citenamefont {Chen}, \citenamefont {Lu},\ and\ \citenamefont {Chen}}]{7.3Gong2010AllOW}%
  \BibitemOpen
  \bibfield  {author} {\bibinfo {author} {\bibfnamefont {M.}~\bibnamefont {Gong}}, \bibinfo {author} {\bibfnamefont {Y.}~\bibnamefont {Chen}}, \bibinfo {author} {\bibfnamefont {F.}~\bibnamefont {Lu}},\ and\ \bibinfo {author} {\bibfnamefont {X.}~\bibnamefont {Chen}},\ }\href {https://api.semanticscholar.org/CorpusID:207340133} {\bibfield  {journal} {\bibinfo  {journal} {Optics letters}\ }\textbf {\bibinfo {volume} {35 16}},\ \bibinfo {pages} {2672} (\bibinfo {year} {2010})}\BibitemShut {NoStop}%
\bibitem [{\citenamefont {Qi}\ and\ \citenamefont {Li}(2020)}]{8Qi2020IntegratedLN}%
  \BibitemOpen
  \bibfield  {author} {\bibinfo {author} {\bibfnamefont {Y.}~\bibnamefont {Qi}}\ and\ \bibinfo {author} {\bibfnamefont {Y.}~\bibnamefont {Li}},\ }\href {https://api.semanticscholar.org/CorpusID:219012388} {\bibfield  {journal} {\bibinfo  {journal} {Nanophotonics}\ }\textbf {\bibinfo {volume} {9}},\ \bibinfo {pages} {1287 } (\bibinfo {year} {2020})}\BibitemShut {NoStop}%
\bibitem [{\citenamefont {Lin}\ \emph {et~al.}(2020)\citenamefont {Lin}, \citenamefont {Bo}, \citenamefont {Cheng},\ and\ \citenamefont {Xu}}]{9Lin2020AdvancesIO}%
  \BibitemOpen
  \bibfield  {author} {\bibinfo {author} {\bibfnamefont {J.}~\bibnamefont {Lin}}, \bibinfo {author} {\bibfnamefont {F.}~\bibnamefont {Bo}}, \bibinfo {author} {\bibfnamefont {Y.}~\bibnamefont {Cheng}},\ and\ \bibinfo {author} {\bibfnamefont {J.}~\bibnamefont {Xu}},\ }\href {https://api.semanticscholar.org/CorpusID:225317128} {\bibfield  {journal} {\bibinfo  {journal} {Photonics Research}\ } (\bibinfo {year} {2020})}\BibitemShut {NoStop}%
\bibitem [{\citenamefont {Boes}\ \emph {et~al.}(2023)\citenamefont {Boes}, \citenamefont {Chang}, \citenamefont {Langrock}, \citenamefont {Yu}, \citenamefont {Zhang}, \citenamefont {Lin}, \citenamefont {Lon$\breve{c}$ar}, \citenamefont {Fejer}, \citenamefont {Bowers},\ and\ \citenamefont {Mitchell}}]{10Boes2023LithiumNP}%
  \BibitemOpen
  \bibfield  {author} {\bibinfo {author} {\bibfnamefont {A.}~\bibnamefont {Boes}}, \bibinfo {author} {\bibfnamefont {L.}~\bibnamefont {Chang}}, \bibinfo {author} {\bibfnamefont {C.}~\bibnamefont {Langrock}}, \bibinfo {author} {\bibfnamefont {M.}~\bibnamefont {Yu}}, \bibinfo {author} {\bibfnamefont {M.}~\bibnamefont {Zhang}}, \bibinfo {author} {\bibfnamefont {Q.}~\bibnamefont {Lin}}, \bibinfo {author} {\bibfnamefont {M.}~\bibnamefont {Lon$\breve{c}$ar}}, \bibinfo {author} {\bibfnamefont {M.}~\bibnamefont {Fejer}}, \bibinfo {author} {\bibfnamefont {J.~E.}\ \bibnamefont {Bowers}},\ and\ \bibinfo {author} {\bibfnamefont {A.}~\bibnamefont {Mitchell}},\ }\href {https://api.semanticscholar.org/CorpusID:255440995} {\bibfield  {journal} {\bibinfo  {journal} {Science}\ }\textbf {\bibinfo {volume} {379}} (\bibinfo {year} {2023})}\BibitemShut {NoStop}%
\bibitem [{\citenamefont {Ma}\ \emph {et~al.}(2020)\citenamefont {Ma}, \citenamefont {Chen}, \citenamefont {Li}, \citenamefont {Tang}, \citenamefont {Sua}, \citenamefont {Fan},\ and\ \citenamefont {Huang}}]{ma2020ultrabright}%
  \BibitemOpen
  \bibfield  {author} {\bibinfo {author} {\bibfnamefont {Z.}~\bibnamefont {Ma}}, \bibinfo {author} {\bibfnamefont {J.-Y.}\ \bibnamefont {Chen}}, \bibinfo {author} {\bibfnamefont {Z.}~\bibnamefont {Li}}, \bibinfo {author} {\bibfnamefont {C.}~\bibnamefont {Tang}}, \bibinfo {author} {\bibfnamefont {Y.~M.}\ \bibnamefont {Sua}}, \bibinfo {author} {\bibfnamefont {H.}~\bibnamefont {Fan}},\ and\ \bibinfo {author} {\bibfnamefont {Y.-P.}\ \bibnamefont {Huang}},\ }\href@noop {} {\bibfield  {journal} {\bibinfo  {journal} {Physical Review Letters}\ }\textbf {\bibinfo {volume} {125}},\ \bibinfo {pages} {263602} (\bibinfo {year} {2020})}\BibitemShut {NoStop}%
\bibitem [{\citenamefont {Chen}\ \emph {et~al.}(2021)\citenamefont {Chen}, \citenamefont {Li}, \citenamefont {Ma}, \citenamefont {Tang}, \citenamefont {Fan}, \citenamefont {Sua},\ and\ \citenamefont {Huang}}]{chen2021photon}%
  \BibitemOpen
  \bibfield  {author} {\bibinfo {author} {\bibfnamefont {J.-Y.}\ \bibnamefont {Chen}}, \bibinfo {author} {\bibfnamefont {Z.}~\bibnamefont {Li}}, \bibinfo {author} {\bibfnamefont {Z.}~\bibnamefont {Ma}}, \bibinfo {author} {\bibfnamefont {C.}~\bibnamefont {Tang}}, \bibinfo {author} {\bibfnamefont {H.}~\bibnamefont {Fan}}, \bibinfo {author} {\bibfnamefont {Y.~M.}\ \bibnamefont {Sua}},\ and\ \bibinfo {author} {\bibfnamefont {Y.-P.}\ \bibnamefont {Huang}},\ }\href@noop {} {\bibfield  {journal} {\bibinfo  {journal} {Physical Review Applied}\ }\textbf {\bibinfo {volume} {16}},\ \bibinfo {pages} {064004} (\bibinfo {year} {2021})}\BibitemShut {NoStop}
\bibitem [{\citenamefont {Wang}\ \emph {et~al.}(2018{\natexlab{a}})\citenamefont {Wang}, \citenamefont {Langrock}, \citenamefont {Marandi}, \citenamefont {Jankowski}, \citenamefont {Zhang}, \citenamefont {Desiatov}, \citenamefont {Fejer},\ and\ \citenamefont {Lon$\breve{c}$ar}}]{11Wang2018UltrahighefficiencyWC}%
  \BibitemOpen
  \bibfield  {author} {\bibinfo {author} {\bibfnamefont {C.}~\bibnamefont {Wang}}, \bibinfo {author} {\bibfnamefont {C.}~\bibnamefont {Langrock}}, \bibinfo {author} {\bibfnamefont {A.}~\bibnamefont {Marandi}}, \bibinfo {author} {\bibfnamefont {M.}~\bibnamefont {Jankowski}}, \bibinfo {author} {\bibfnamefont {M.}~\bibnamefont {Zhang}}, \bibinfo {author} {\bibfnamefont {B.}~\bibnamefont {Desiatov}}, \bibinfo {author} {\bibfnamefont {M.~M.}\ \bibnamefont {Fejer}},\ and\ \bibinfo {author} {\bibfnamefont {M.}~\bibnamefont {Lon$\breve{c}$ar}},\ }\href {https://api.semanticscholar.org/CorpusID:96448072} {\bibfield  {journal} {\bibinfo  {journal} {Optica}\ } (\bibinfo {year} {2018}{\natexlab{a}})}\BibitemShut {NoStop}%
\bibitem [{\citenamefont {yang Chen}\ \emph {et~al.}(2019{\natexlab{a}})\citenamefont {yang Chen}, \citenamefont {Sua}, \citenamefont {Ma}, \citenamefont {Tang}, \citenamefont {Li},\ and\ \citenamefont {Huang}}]{12Chen2019EfficientPF}%
  \BibitemOpen
  \bibfield  {author} {\bibinfo {author} {\bibfnamefont {J.}~\bibnamefont {yang Chen}}, \bibinfo {author} {\bibfnamefont {Y.~M.}\ \bibnamefont {Sua}}, \bibinfo {author} {\bibfnamefont {Z.}~\bibnamefont {Ma}}, \bibinfo {author} {\bibfnamefont {C.}~\bibnamefont {Tang}}, \bibinfo {author} {\bibfnamefont {Z.}~\bibnamefont {Li}},\ and\ \bibinfo {author} {\bibfnamefont {Y.-P.}\ \bibnamefont {Huang}},\ }\href {https://api.semanticscholar.org/CorpusID:84842683} {\bibfield  {journal} {\bibinfo  {journal} {OSA Continuum}\ } (\bibinfo {year} {2019}{\natexlab{a}})}\BibitemShut {NoStop}%
\bibitem [{\citenamefont {Zhao}\ \emph {et~al.}(2020)\citenamefont {Zhao}, \citenamefont {R{\"u}sing}, \citenamefont {Javid}, \citenamefont {Ling}, \citenamefont {Li}, \citenamefont {Lin},\ and\ \citenamefont {Mookherjea}}]{13Zhao2020ShallowetchedTL}%
  \BibitemOpen
  \bibfield  {author} {\bibinfo {author} {\bibfnamefont {J.}~\bibnamefont {Zhao}}, \bibinfo {author} {\bibfnamefont {M.}~\bibnamefont {R{\"u}sing}}, \bibinfo {author} {\bibfnamefont {U.~A.}\ \bibnamefont {Javid}}, \bibinfo {author} {\bibfnamefont {J.}~\bibnamefont {Ling}}, \bibinfo {author} {\bibfnamefont {M.}~\bibnamefont {Li}}, \bibinfo {author} {\bibfnamefont {Q.}~\bibnamefont {Lin}},\ and\ \bibinfo {author} {\bibfnamefont {S.}~\bibnamefont {Mookherjea}},\ }\href {https://api.semanticscholar.org/CorpusID:220586453} {\bibfield  {journal} {\bibinfo  {journal} {Optics express}\ }\textbf {\bibinfo {volume} {28 13}},\ \bibinfo {pages} {19669} (\bibinfo {year} {2020})}\BibitemShut {NoStop}%
\bibitem [{\citenamefont {Wang}\ \emph {et~al.}(2018{\natexlab{b}})\citenamefont {Wang}, \citenamefont {Zhang}, \citenamefont {Yu}, \citenamefont {Zhu}, \citenamefont {Hu},\ and\ \citenamefont {Lon$\breve{c}$ar}}]{14Wang2018MonolithicLN}%
  \BibitemOpen
  \bibfield  {author} {\bibinfo {author} {\bibfnamefont {C.}~\bibnamefont {Wang}}, \bibinfo {author} {\bibfnamefont {M.}~\bibnamefont {Zhang}}, \bibinfo {author} {\bibfnamefont {M.}~\bibnamefont {Yu}}, \bibinfo {author} {\bibfnamefont {R.}~\bibnamefont {Zhu}}, \bibinfo {author} {\bibfnamefont {H.}~\bibnamefont {Hu}},\ and\ \bibinfo {author} {\bibfnamefont {M.}~\bibnamefont {Lon$\breve{c}$ar}},\ }\href {https://api.semanticscholar.org/CorpusID:52904678} {\bibfield  {journal} {\bibinfo  {journal} {Nature Communications}\ }\textbf {\bibinfo {volume} {10}} (\bibinfo {year} {2018}{\natexlab{b}})}\BibitemShut {NoStop}%
\bibitem [{\citenamefont {Ge}\ \emph {et~al.}(2018)\citenamefont {Ge}, \citenamefont {Chen}, \citenamefont {Jiang}, \citenamefont {Li}, \citenamefont {Zhu}, \citenamefont {Liu},\ and\ \citenamefont {Chen}}]{15Ge2018BroadbandQM}%
  \BibitemOpen
  \bibfield  {author} {\bibinfo {author} {\bibfnamefont {L.}~\bibnamefont {Ge}}, \bibinfo {author} {\bibfnamefont {Y.}~\bibnamefont {Chen}}, \bibinfo {author} {\bibfnamefont {H.}~\bibnamefont {Jiang}}, \bibinfo {author} {\bibfnamefont {G.}~\bibnamefont {Li}}, \bibinfo {author} {\bibfnamefont {B.}~\bibnamefont {Zhu}}, \bibinfo {author} {\bibfnamefont {Y.}~\bibnamefont {Liu}},\ and\ \bibinfo {author} {\bibfnamefont {X.}~\bibnamefont {Chen}},\ }\href@noop {} {\bibfield  {journal} {\bibinfo  {journal} {Photonics Research}\ } (\bibinfo {year} {2018})}\BibitemShut {NoStop}%
\bibitem [{\citenamefont {Jankowski}\ \emph {et~al.}(2019)\citenamefont {Jankowski}, \citenamefont {Langrock}, \citenamefont {Desiatov}, \citenamefont {Marandi}, \citenamefont {Wang}, \citenamefont {Zhang}, \citenamefont {Phillips}, \citenamefont {Lon$\breve{c}$ar},\ and\ \citenamefont {Fejer}}]{16Jankowski2019UltrabroadbandNO}%
  \BibitemOpen
  \bibfield  {author} {\bibinfo {author} {\bibfnamefont {M.}~\bibnamefont {Jankowski}}, \bibinfo {author} {\bibfnamefont {C.}~\bibnamefont {Langrock}}, \bibinfo {author} {\bibfnamefont {B.}~\bibnamefont {Desiatov}}, \bibinfo {author} {\bibfnamefont {A.}~\bibnamefont {Marandi}}, \bibinfo {author} {\bibfnamefont {C.}~\bibnamefont {Wang}}, \bibinfo {author} {\bibfnamefont {M.}~\bibnamefont {Zhang}}, \bibinfo {author} {\bibfnamefont {C.~R.}\ \bibnamefont {Phillips}}, \bibinfo {author} {\bibfnamefont {M.}~\bibnamefont {Lon$\breve{c}$ar}},\ and\ \bibinfo {author} {\bibfnamefont {M.~M.}\ \bibnamefont {Fejer}},\ }\href {https://api.semanticscholar.org/CorpusID:202677653} {\bibfield  {journal} {\bibinfo  {journal} {Optica}\ } (\bibinfo {year} {2019})}\BibitemShut {NoStop}%
\bibitem [{\citenamefont {Mishra}\ \emph {et~al.}(2022)\citenamefont {Mishra}, \citenamefont {Jankowski}, \citenamefont {Hwang}, \citenamefont {Stokowski}, \citenamefont {McKenna}, \citenamefont {Langrock}, \citenamefont {Ng}, \citenamefont {Heydari}, \citenamefont {Mabuchi}, \citenamefont {Safavi-Naeini},\ and\ \citenamefont {Fejer}}]{17Mishra:22}%
  \BibitemOpen
  \bibfield  {author} {\bibinfo {author} {\bibfnamefont {J.}~\bibnamefont {Mishra}}, \bibinfo {author} {\bibfnamefont {M.}~\bibnamefont {Jankowski}}, \bibinfo {author} {\bibfnamefont {A.~Y.}\ \bibnamefont {Hwang}}, \bibinfo {author} {\bibfnamefont {H.~S.}\ \bibnamefont {Stokowski}}, \bibinfo {author} {\bibfnamefont {T.~P.}\ \bibnamefont {McKenna}}, \bibinfo {author} {\bibfnamefont {C.}~\bibnamefont {Langrock}}, \bibinfo {author} {\bibfnamefont {E.}~\bibnamefont {Ng}}, \bibinfo {author} {\bibfnamefont {D.}~\bibnamefont {Heydari}}, \bibinfo {author} {\bibfnamefont {H.}~\bibnamefont {Mabuchi}}, \bibinfo {author} {\bibfnamefont {A.~H.}\ \bibnamefont {Safavi-Naeini}},\ and\ \bibinfo {author} {\bibfnamefont {M.~M.}\ \bibnamefont {Fejer}},\ }\href {https://doi.org/10.1364/OE.467580} {\bibfield  {journal} {\bibinfo  {journal} {Opt. Express}\ }\textbf {\bibinfo {volume} {30}},\ \bibinfo {pages} {32752} (\bibinfo {year} {2022})}\BibitemShut {NoStop}%
\bibitem [{\citenamefont {Javid}\ \emph {et~al.}(2021{\natexlab{b}})\citenamefont {Javid}, \citenamefont {Ling}, \citenamefont {Staffa}, \citenamefont {Li}, \citenamefont {He},\ and\ \citenamefont {Lin}}]{18Javid2021UltrabroadbandEP}%
  \BibitemOpen
  \bibfield  {author} {\bibinfo {author} {\bibfnamefont {U.~A.}\ \bibnamefont {Javid}}, \bibinfo {author} {\bibfnamefont {J.}~\bibnamefont {Ling}}, \bibinfo {author} {\bibfnamefont {J.}~\bibnamefont {Staffa}}, \bibinfo {author} {\bibfnamefont {M.}~\bibnamefont {Li}}, \bibinfo {author} {\bibfnamefont {Y.}~\bibnamefont {He}},\ and\ \bibinfo {author} {\bibfnamefont {Q.}~\bibnamefont {Lin}},\ }\href {https://api.semanticscholar.org/CorpusID:231592825} {\bibfield  {journal} {\bibinfo  {journal} {Physical review letters}\ }\textbf {\bibinfo {volume} {127 18}},\ \bibinfo {pages} {183601} (\bibinfo {year} {2021}{\natexlab{b}})}\BibitemShut {NoStop}%
\bibitem [{\citenamefont {ning Zhu}\ \emph {et~al.}(1997)\citenamefont {ning Zhu}, \citenamefont {Zhu},\ and\ \citenamefont {Ming}}]{19Zhu1997QuasiphasematchedTG}%
  \BibitemOpen
  \bibfield  {author} {\bibinfo {author} {\bibfnamefont {S.}~\bibnamefont {ning Zhu}}, \bibinfo {author} {\bibfnamefont {Y.}~\bibnamefont {Zhu}},\ and\ \bibinfo {author} {\bibfnamefont {N.}~\bibnamefont {Ming}},\ }\href {https://api.semanticscholar.org/CorpusID:110932807} {\bibfield  {journal} {\bibinfo  {journal} {Science}\ }\textbf {\bibinfo {volume} {278}},\ \bibinfo {pages} {843} (\bibinfo {year} {1997})}\BibitemShut {NoStop}%
\bibitem [{\citenamefont {Paul}\ \emph {et~al.}(2003)\citenamefont {Paul}, \citenamefont {Bartels}, \citenamefont {Tobey}, \citenamefont {Green}, \citenamefont {Weiman}, \citenamefont {Murnane}, \citenamefont {Kapteyn}, \citenamefont {Backus},\ and\ \citenamefont {Christov}}]{20Paul2003QuasiphasematchedGO}%
  \BibitemOpen
  \bibfield  {author} {\bibinfo {author} {\bibfnamefont {A.}~\bibnamefont {Paul}}, \bibinfo {author} {\bibfnamefont {R.~A.}\ \bibnamefont {Bartels}}, \bibinfo {author} {\bibfnamefont {R.~I.}\ \bibnamefont {Tobey}}, \bibinfo {author} {\bibfnamefont {H.}~\bibnamefont {Green}}, \bibinfo {author} {\bibfnamefont {S.}~\bibnamefont {Weiman}}, \bibinfo {author} {\bibfnamefont {M.~M.}\ \bibnamefont {Murnane}}, \bibinfo {author} {\bibfnamefont {H.~C.}\ \bibnamefont {Kapteyn}}, \bibinfo {author} {\bibfnamefont {S.~J.}\ \bibnamefont {Backus}},\ and\ \bibinfo {author} {\bibfnamefont {I.~P.}\ \bibnamefont {Christov}},\ }\href {https://api.semanticscholar.org/CorpusID:4420103} {\bibfield  {journal} {\bibinfo  {journal} {Nature}\ }\textbf {\bibinfo {volume} {421}},\ \bibinfo {pages} {51} (\bibinfo {year} {2003})}\BibitemShut {NoStop}%
\bibitem [{\citenamefont {Chen}\ \emph {et~al.}(2015)\citenamefont {Chen}, \citenamefont {Zhang}, \citenamefont {Hu}, \citenamefont {Liu},\ and\ \citenamefont {Li}}]{21Chen2015HighEfficiencyBH}%
  \BibitemOpen
  \bibfield  {author} {\bibinfo {author} {\bibfnamefont {B.-Q.}\ \bibnamefont {Chen}}, \bibinfo {author} {\bibfnamefont {C.}~\bibnamefont {Zhang}}, \bibinfo {author} {\bibfnamefont {C.-Y.}\ \bibnamefont {Hu}}, \bibinfo {author} {\bibfnamefont {R.-J.}\ \bibnamefont {Liu}},\ and\ \bibinfo {author} {\bibfnamefont {Z.-Y.}\ \bibnamefont {Li}},\ }\href {https://api.semanticscholar.org/CorpusID:17098198} {\bibfield  {journal} {\bibinfo  {journal} {Physical review letters}\ }\textbf {\bibinfo {volume} {115 8}},\ \bibinfo {pages} {083902} (\bibinfo {year} {2015})}\BibitemShut {NoStop}%
\bibitem [{\citenamefont {Huang}\ \emph {et~al.}(2022)\citenamefont {Huang}, \citenamefont {Fang}, \citenamefont {Yan}, \citenamefont {Wu},\ and\ \citenamefont {ping Zeng}}]{22Huang2022WidefieldMS}%
  \BibitemOpen
  \bibfield  {author} {\bibinfo {author} {\bibfnamefont {K.}~\bibnamefont {Huang}}, \bibinfo {author} {\bibfnamefont {J.}~\bibnamefont {Fang}}, \bibinfo {author} {\bibfnamefont {M.}~\bibnamefont {Yan}}, \bibinfo {author} {\bibfnamefont {E.}~\bibnamefont {Wu}},\ and\ \bibinfo {author} {\bibfnamefont {H.}~\bibnamefont {ping Zeng}},\ }\href {https://api.semanticscholar.org/CorpusID:247168781} {\bibfield  {journal} {\bibinfo  {journal} {Nature Communications}\ }\textbf {\bibinfo {volume} {13}} (\bibinfo {year} {2022})}\BibitemShut {NoStop}%
\bibitem [{\citenamefont {Wang}\ \emph {et~al.}(2017)\citenamefont {Wang}, \citenamefont {Li}, \citenamefont {hwan Kim}, \citenamefont {Xiong}, \citenamefont {Ren}, \citenamefont {Guo}, \citenamefont {Yu},\ and\ \citenamefont {Lon$\breve{c}$ar}}]{23Wang2017MetasurfaceassistedPS}%
  \BibitemOpen
  \bibfield  {author} {\bibinfo {author} {\bibfnamefont {C.}~\bibnamefont {Wang}}, \bibinfo {author} {\bibfnamefont {Z.}~\bibnamefont {Li}}, \bibinfo {author} {\bibfnamefont {M.}~\bibnamefont {hwan Kim}}, \bibinfo {author} {\bibfnamefont {X.}~\bibnamefont {Xiong}}, \bibinfo {author} {\bibfnamefont {X.}~\bibnamefont {Ren}}, \bibinfo {author} {\bibfnamefont {G.}~\bibnamefont {Guo}}, \bibinfo {author} {\bibfnamefont {N.}~\bibnamefont {Yu}},\ and\ \bibinfo {author} {\bibfnamefont {M.}~\bibnamefont {Lon$\breve{c}$ar}},\ }\href {https://api.semanticscholar.org/CorpusID:527391} {\bibfield  {journal} {\bibinfo  {journal} {Nature Communications}\ }\textbf {\bibinfo {volume} {8}} (\bibinfo {year} {2017})}\BibitemShut {NoStop}%
\bibitem [{\citenamefont {Lin}\ \emph {et~al.}(2017)\citenamefont {Lin}, \citenamefont {Coillet},\ and\ \citenamefont {Chembo}}]{24Lin2017NonlinearPW}%
  \BibitemOpen
  \bibfield  {author} {\bibinfo {author} {\bibfnamefont {J.}~\bibnamefont {yang Chen}}, \bibinfo {author} {\bibfnamefont {Z.}~\bibnamefont {Ma}}, \bibinfo {author} {\bibfnamefont {Y.~M.}\ \bibnamefont {Sua}}, \bibinfo {author} {\bibfnamefont {Z.}~\bibnamefont {Li}}, \bibinfo {author} {\bibfnamefont {C.}~\bibnamefont {Tang}},\ and\ \bibinfo {author} {\bibfnamefont {Y.-P.}\ \bibnamefont {Huang}},\ }\href {https://api.semanticscholar.org/CorpusID:204292904} {\bibfield  {journal} {\bibinfo  {journal} {Optica}\ } (\bibinfo {year} {2019}{\natexlab{b}})}\BibitemShut {NoStop}%
\bibitem [{\citenamefont {Huang}\ \emph {et~al.}(2004)\citenamefont {Huang}, \citenamefont {Kurz}, \citenamefont {Langrock}, \citenamefont {Schober},\ and\ \citenamefont {Fejer}}]{38Huang2004QuasigroupvelocityMU}%
  \BibitemOpen
  \bibfield  {author} {\bibinfo {author} {\bibfnamefont {J.}~\bibnamefont {Huang}}, \bibinfo {author} {\bibfnamefont {J.~R.}\ \bibnamefont {Kurz}}, \bibinfo {author} {\bibfnamefont {C.}~\bibnamefont {Langrock}}, \bibinfo {author} {\bibfnamefont {A.}~\bibnamefont {Schober}},\ and\ \bibinfo {author} {\bibfnamefont {M.~M.}\ \bibnamefont {Fejer}},\ }\href@noop {} {\bibfield  {journal} {\bibinfo  {journal} {Optics letters}\ }\textbf {\bibinfo {volume} {29 21}},\ \bibinfo {pages} {2482} (\bibinfo {year} {2004})}\BibitemShut {NoStop}%
\bibitem [{\citenamefont {Xie}\ \emph {et~al.}(2006)\citenamefont {Xie}, \citenamefont {Huang},\ and\ \citenamefont {Fejer}}]{39Xie2006NarrowlinewidthNO}%
  \BibitemOpen
  \bibfield  {author} {\bibinfo {author} {\bibfnamefont {X.}~\bibnamefont {Xie}}, \bibinfo {author} {\bibfnamefont {J.}~\bibnamefont {Huang}},\ and\ \bibinfo {author} {\bibfnamefont {M.~M.}\ \bibnamefont {Fejer}},\ }\href {https://api.semanticscholar.org/CorpusID:21967568} {\bibfield  {journal} {\bibinfo  {journal} {Optics letters}\ }\textbf {\bibinfo {volume} {31 14}},\ \bibinfo {pages} {2190} (\bibinfo {year} {2006})}\BibitemShut {NoStop}%
\bibitem [{\citenamefont {Mao}\ \emph {et~al.}(2021)\citenamefont {Mao}, \citenamefont {Wang}, \citenamefont {Zhang}, \citenamefont {Zeng}, \citenamefont {Du}, \citenamefont {He}, \citenamefont {Sun},\ and\ \citenamefont {Zhao}}]{40Mao2021SynchronizedMS}%
  \BibitemOpen
  \bibfield  {author} {\bibinfo {author} {\bibfnamefont {D.}~\bibnamefont {Mao}}, \bibinfo {author} {\bibfnamefont {H.}~\bibnamefont {Wang}}, \bibinfo {author} {\bibfnamefont {H.}~\bibnamefont {Zhang}}, \bibinfo {author} {\bibfnamefont {C.}~\bibnamefont {Zeng}}, \bibinfo {author} {\bibfnamefont {Y.}~\bibnamefont {Du}}, \bibinfo {author} {\bibfnamefont {Z.}~\bibnamefont {He}}, \bibinfo {author} {\bibfnamefont {Z.}~\bibnamefont {Sun}},\ and\ \bibinfo {author} {\bibfnamefont {J.}~\bibnamefont {Zhao}},\ }\href {https://api.semanticscholar.org/CorpusID:244403180} {\bibfield  {journal} {\bibinfo  {journal} {Nature Communications}\ }\textbf {\bibinfo {volume} {12}} (\bibinfo {year} {2021})}\BibitemShut {NoStop}%
\bibitem [{\citenamefont {Lourdesamy}\ \emph {et~al.}(2021)\citenamefont {Lourdesamy}, \citenamefont {Runge}, \citenamefont {Alexander}, \citenamefont {Hudson}, \citenamefont {Blanco-Redondo},\ and\ \citenamefont {de~Sterke}}]{41Lourdesamy2021SpectrallyPP}%
  \BibitemOpen
  \bibfield  {author} {\bibinfo {author} {\bibfnamefont {J.~P.}\ \bibnamefont {Lourdesamy}}, \bibinfo {author} {\bibfnamefont {A.~F.~J.}\ \bibnamefont {Runge}}, \bibinfo {author} {\bibfnamefont {T.~J.}\ \bibnamefont {Alexander}}, \bibinfo {author} {\bibfnamefont {D.~D.}\ \bibnamefont {Hudson}}, \bibinfo {author} {\bibfnamefont {A.}~\bibnamefont {Blanco-Redondo}},\ and\ \bibinfo {author} {\bibfnamefont {C.~M.}\ \bibnamefont {de~Sterke}},\ }\href {https://api.semanticscholar.org/CorpusID:245271342} {\bibfield  {journal} {\bibinfo  {journal} {Nature Physics}\ }\textbf {\bibinfo {volume} {18}},\ \bibinfo {pages} {59 } (\bibinfo {year} {2021})}\BibitemShut {NoStop}%
\bibitem [{\citenamefont {Cui}\ \emph {et~al.}(2023)\citenamefont {Cui}, \citenamefont {Zhang}, \citenamefont {Huang}, \citenamefont {Zhang}, \citenamefont {Liu}, \citenamefont {Kuang}, \citenamefont {Tao}, \citenamefont {Chen}, \citenamefont {Liu},\ and\ \citenamefont {Malomed}}]{42Cui12022DichromaticM}%
  \BibitemOpen
  \bibfield  {author} {\bibinfo {author} {\bibfnamefont {M.}~\bibnamefont {Heiblum}}\ and\ \bibinfo {author} {\bibfnamefont {J.~H.}\ \bibnamefont {Harris}},\ }\href@noop {} {\bibfield  {journal} {\bibinfo  {journal} {IEEE Journal of Quantum Electronics}\ }\textbf {\bibinfo {volume} {11}},\ \bibinfo {pages} {75} (\bibinfo {year} {1975})}\BibitemShut {NoStop}%
\bibitem [{\citenamefont {Wang}\ \emph {et~al.}(2018)\citenamefont {Wang}, \citenamefont {Zhang}, \citenamefont {Chen}, \citenamefont {Bertrand}, \citenamefont {Shams-Ansari}, \citenamefont {Chandrasekhar}, \citenamefont {Winzer},\ and\ \citenamefont {Lon{\v{c}}ar}}]{wang2018integrated}%
  \BibitemOpen
  \bibfield  {author} {\bibinfo {author} {\bibfnamefont {C.}~\bibnamefont {Wang}}, \bibinfo {author} {\bibfnamefont {M.}~\bibnamefont {Zhang}}, \bibinfo {author} {\bibfnamefont {X.}~\bibnamefont {Chen}}, \bibinfo {author} {\bibfnamefont {M.}~\bibnamefont {Bertrand}}, \bibinfo {author} {\bibfnamefont {A.}~\bibnamefont {Shams-Ansari}}, \bibinfo {author} {\bibfnamefont {S.}~\bibnamefont {Chandrasekhar}}, \bibinfo {author} {\bibfnamefont {P.}~\bibnamefont {Winzer}},\ and\ \bibinfo {author} {\bibfnamefont {M.}~\bibnamefont {Lon{\v{c}}ar}},\ }\href@noop {} {\bibfield  {journal} {\bibinfo  {journal} {Nature}\ }\textbf {\bibinfo {volume} {562}},\ \bibinfo {pages} {101} (\bibinfo {year} {2018})}\BibitemShut {NoStop}%
\bibitem [{\citenamefont {Zhang}\ \emph {et~al.}(2019)\citenamefont {Zhang}, \citenamefont {Buscaino}, \citenamefont {Wang}, \citenamefont {Shams-Ansari}, \citenamefont {Reimer}, \citenamefont {Zhu}, \citenamefont {Kahn},\ and\ \citenamefont {Lon{\v{c}}ar}}]{zhang2019broadband}%
  \BibitemOpen
  \bibfield  {author} {\bibinfo {author} {\bibfnamefont {M.}~\bibnamefont {Zhang}}, \bibinfo {author} {\bibfnamefont {B.}~\bibnamefont {Buscaino}}, \bibinfo {author} {\bibfnamefont {C.}~\bibnamefont {Wang}}, \bibinfo {author} {\bibfnamefont {A.}~\bibnamefont {Shams-Ansari}}, \bibinfo {author} {\bibfnamefont {C.}~\bibnamefont {Reimer}}, \bibinfo {author} {\bibfnamefont {R.}~\bibnamefont {Zhu}}, \bibinfo {author} {\bibfnamefont {J.~M.}\ \bibnamefont {Kahn}},\ and\ \bibinfo {author} {\bibfnamefont {M.}~\bibnamefont {Lon{\v{c}}ar}},\ }\href@noop {} {\bibfield  {journal} {\bibinfo  {journal} {Nature}\ }\textbf {\bibinfo {volume} {568}},\ \bibinfo {pages} {373} (\bibinfo {year} {2019})}\BibitemShut {NoStop}%
\bibitem [{\citenamefont {Han}\ \emph {et~al.}(2022)\citenamefont {Han}, \citenamefont {Jiang}, \citenamefont {Frigg}, \citenamefont {Xiao}, \citenamefont {Zhang}, \citenamefont {Nguyen}, \citenamefont {Boes}, \citenamefont {Yang}, \citenamefont {Ren}, \citenamefont {Su} \emph {et~al.}}]{han2022mode}%
  \BibitemOpen
  \bibfield  {author} {\bibinfo {author} {\bibfnamefont {X.}~\bibnamefont {Han}}, \bibinfo {author} {\bibfnamefont {Y.}~\bibnamefont {Jiang}}, \bibinfo {author} {\bibfnamefont {A.}~\bibnamefont {Frigg}}, \bibinfo {author} {\bibfnamefont {H.}~\bibnamefont {Xiao}}, \bibinfo {author} {\bibfnamefont {P.}~\bibnamefont {Zhang}}, \bibinfo {author} {\bibfnamefont {T.~G.}\ \bibnamefont {Nguyen}}, \bibinfo {author} {\bibfnamefont {A.}~\bibnamefont {Boes}}, \bibinfo {author} {\bibfnamefont {J.}~\bibnamefont {Yang}}, \bibinfo {author} {\bibfnamefont {G.}~\bibnamefont {Ren}}, \bibinfo {author} {\bibfnamefont {Y.}~\bibnamefont {Su}}, \emph {et~al.},\ }\href@noop {} {\bibfield  {journal} {\bibinfo  {journal} {Laser \& Photonics Reviews}\ }\textbf {\bibinfo {volume} {16}},\ \bibinfo {pages} {2100529} (\bibinfo {year} {2022})}\BibitemShut {NoStop}%
\bibitem [{\citenamefont {Wang}\ \emph {et~al.}(2020)\citenamefont {Wang}, \citenamefont {Chen}, \citenamefont {Dai},\ and\ \citenamefont {Liu}}]{9095209}%
  \BibitemOpen
  \bibfield  {author} {\bibinfo {author} {\bibfnamefont {J.}~\bibnamefont {Wang}}, \bibinfo {author} {\bibfnamefont {P.}~\bibnamefont {Chen}}, \bibinfo {author} {\bibfnamefont {D.}~\bibnamefont {Dai}},\ and\ \bibinfo {author} {\bibfnamefont {L.}~\bibnamefont {Liu}},\ }\href {https://doi.org/10.1109/JPHOT.2020.2995317} {\bibfield  {journal} {\bibinfo  {journal} {IEEE Photonics Journal}\ }\textbf {\bibinfo {volume} {12}},\ \bibinfo {pages} {1} (\bibinfo {year} {2020})}\BibitemShut {NoStop}%
\bibitem [{\citenamefont {Wang}\ \emph {et~al.}(2024)\citenamefont {Wang}, \citenamefont {Yuan}, \citenamefont {Wu}, \citenamefont {Chen},\ and\ \citenamefont {Chen}}]{wang2024enhanced}%
  \BibitemOpen
  \bibfield  {author} {\bibinfo {author} {\bibfnamefont {X.}~\bibnamefont {Wang}}, \bibinfo {author} {\bibfnamefont {T.}~\bibnamefont {Yuan}}, \bibinfo {author} {\bibfnamefont {J.}~\bibnamefont {Wu}}, \bibinfo {author} {\bibfnamefont {Y.}~\bibnamefont {Chen}},\ and\ \bibinfo {author} {\bibfnamefont {X.}~\bibnamefont {Chen}},\ }\href@noop {} {\bibfield  {journal} {\bibinfo  {journal} {Laser \& Photonics Reviews}\ ,\ \bibinfo {pages} {2300760}} (\bibinfo {year} {2024})}\BibitemShut {NoStop}%
\bibitem [{\citenamefont {Lin}\ \emph {et~al.}(2019)\citenamefont {Lin}, \citenamefont {Yao}, \citenamefont {Hao}, \citenamefont {Zhang}, \citenamefont {Mao}, \citenamefont {Wang}, \citenamefont {Chu}, \citenamefont {Wu}, \citenamefont {Fang}, \citenamefont {Qiao} \emph {et~al.}}]{lin2019broadband}%
  \BibitemOpen
  \bibfield  {author} {\bibinfo {author} {\bibfnamefont {J.}~\bibnamefont {Lin}}, \bibinfo {author} {\bibfnamefont {N.}~\bibnamefont {Yao}}, \bibinfo {author} {\bibfnamefont {Z.}~\bibnamefont {Hao}}, \bibinfo {author} {\bibfnamefont {J.}~\bibnamefont {Zhang}}, \bibinfo {author} {\bibfnamefont {W.}~\bibnamefont {Mao}}, \bibinfo {author} {\bibfnamefont {M.}~\bibnamefont {Wang}}, \bibinfo {author} {\bibfnamefont {W.}~\bibnamefont {Chu}}, \bibinfo {author} {\bibfnamefont {R.}~\bibnamefont {Wu}}, \bibinfo {author} {\bibfnamefont {Z.}~\bibnamefont {Fang}}, \bibinfo {author} {\bibfnamefont {L.}~\bibnamefont {Qiao}}, \emph {et~al.},\ }\href@noop {} {\bibfield  {journal} {\bibinfo  {journal} {Physical review letters}\ }\textbf {\bibinfo {volume} {122}},\ \bibinfo {pages} {173903} (\bibinfo {year} {2019})}\BibitemShut {NoStop}%
\bibitem [{\citenamefont {Yuan}\ \emph {et~al.}(2023)\citenamefont {Yuan}, \citenamefont {Wu}, \citenamefont {Liu}, \citenamefont {Yan}, \citenamefont {Li}, \citenamefont {Jiang}, \citenamefont {Lin}, \citenamefont {Chen},\ and\ \citenamefont {Chen}}]{352023ChipscaleSQ}%
  \BibitemOpen
  \bibfield  {author} {\bibinfo {author} {\bibfnamefont {T.}~\bibnamefont {Yuan}}, \bibinfo {author} {\bibfnamefont {J.}~\bibnamefont {Wu}}, \bibinfo {author} {\bibfnamefont {Y.}~\bibnamefont {Liu}}, \bibinfo {author} {\bibfnamefont {X.}~\bibnamefont {Yan}}, \bibinfo {author} {\bibfnamefont {H.}~\bibnamefont {Li}}, \bibinfo {author} {\bibfnamefont {H.}~\bibnamefont {Jiang}}, \bibinfo {author} {\bibfnamefont {Q.}~\bibnamefont {Lin}}, \bibinfo {author} {\bibfnamefont {Y.}~\bibnamefont {Chen}},\ and\ \bibinfo {author} {\bibfnamefont {X.}~\bibnamefont {Chen}},\ }\href@noop {} {\bibfield  {journal} {\bibinfo  {journal} {Science China Physics, Mechanics \& Astronomy}\ }\textbf {\bibinfo {volume} {66}} (\bibinfo {year} {2023})}\BibitemShut {NoStop}%
\bibitem [{\citenamefont {Pan}\ \emph {et~al.}(2019)\citenamefont {Pan}, \citenamefont {Hu}, \citenamefont {Zeng},\ and\ \citenamefont {Xia}}]{36Pan2019FundamentalMH}%
\BibitemOpen
\bibfield  {author} {\bibinfo {author} {\bibfnamefont {A.}~\bibnamefont {Pan}}, \bibinfo {author} {\bibfnamefont {C.}~\bibnamefont {Hu}}, \bibinfo {author} {\bibfnamefont {C.}~\bibnamefont {Zeng}},\ and\ \bibinfo {author} {\bibfnamefont {J.}~\bibnamefont {Xia}},\ }\href@noop {} {\bibfield  {journal} {\bibinfo  {journal} {Optics express}\ }\textbf {\bibinfo {volume} {27 24}},\ \bibinfo {pages} {35659} (\bibinfo {year} {2019})}\BibitemShut {NoStop}%
\bibitem [{\citenamefont {Liu}\ \emph {et~al.}(2022)\citenamefont {Liu}, \citenamefont {Li}, \citenamefont {Chen},\ and\ \citenamefont {Li}}]{37Liu2022ModalAO}%
  \BibitemOpen
  \bibfield  {author} {\bibinfo {author} {\bibfnamefont {W.-C.}\ \bibnamefont {Liu}}, \bibinfo {author} {\bibfnamefont {Y.}~\bibnamefont {Li}}, \bibinfo {author} {\bibfnamefont {B.}~\bibnamefont {Chen}},\ and\ \bibinfo {author} {\bibfnamefont {Z.}~\bibnamefont {Li}},\ }\href@noop {} {\bibfield  {journal} {\bibinfo  {journal} {Journal of Optics}\ }\textbf {\bibinfo {volume} {24}} (\bibinfo {year} {2022})}\BibitemShut {Stop}%







\BibitemOpen
\end{thebibliography}
\end{document}